\documentstyle[sprocl]{article}

\def\Journal#1#2#3#4{{#1} {\bf #2}, #3 (#4)}

\def\PLB{{\em Phys. Lett.}  B}

\def\PRD{{\em Phys. Rev.} D}

\def\PLB{{\em Phys. Lett.}  B}

\def\PRD{{\em Phys. Rev.} D}

\def\mco{\multicolumn}
\def\epp{\epsilon^{\prime}}
\def\vep{\varepsilon}
\def\ra{\rightarrow}
\def\ppg{\pi^+\pi^-\gamma}
\def\vp{{\bf p}}
\def\ko{K^0}
\def\kb{\bar{K^0}}
\def\al{\alpha}
\def\ab{\bar{\alpha}}
\def\be{\begin{equation}}
\def\ee{\end{equation}}
\def\bea{\begin{eqnarray}}
\def\eea{\end{eqnarray}}
\def\CPbar{\hbox{{\rm CP}\hskip-1.80em{/}}}

\begin{document}

\title{MAXIMALLY SYMMETRIC COSMOLOGICAL SOLUTIONS OF TYPE II SUPERSTRINGS}

\author{ M. C. BENTO\footnote{Talk presented at the New Worlds
 in Astroparticle Physics Conference, Faro, Portugal, September 1998, to be published by World Scientific Press. }}

\address{Departamento de F\'{\i}sica, Instituto Superior T\'ecnico, Av. Rovisco Pais, \\Lisboa Codex, Portugal}




\maketitle\abstracts{
  We study maximally symmetric cosmological solutions of type II supersymmetric strings in the presence of the exact, SL(2,Z)-invariant,  higher-curvature corrections to the lowest order effective action. We find  that, unlike the case of type IIA theories,  de Sitter solutions exist, at all orders in $\alpha'$,  for type IIB superstrings when non-perturbative D-instanton effects are included on the basis of SL(2,Z) invariance.}

Recently, there has been considerable interest on the subject of cosmological solutions of the low-energy limit of string theories and M-theory.
String theories, in their low-energy limit, give rise to effective theories of gravity containing higher-derivative corrections.
The effect of higher-curvature terms in the string  low energy effective actions on the maximally symmetric cosmological solutions of the theory has been studied \cite{bento1}, with the result that, for all string theories, at  tree-level, there are no stable de Sitter solutions and these solutions disappear once the dilaton is included.  However, in those works, instanton and string-loop effects were not taken into account. Here, we report the result of Ref.~[2], where it is shown that maximally symmetric solutions exist for type IIB superstrings, at all orders in string perturbation theory once those effects are taken into account on the basis of SL(2,Z) invariance.

We start by considering the bosonic effective Lagrangian of the theory at lowest order in $\alpha'$ (hereafter we set $\alpha'=1)$ in the Einstein frame 

\begin{equation}
\label{a}
{\cal L}={\cal L}_0+{\cal L}_3,
\end{equation}
where ${\cal L}_0$ is the lowest order Lagrangian and  ${\cal L}_3$ contains the first perturbative  corrections \cite{gross}, which   arise at order  $\alpha'^3$,  
\begin{eqnarray}
\label{aa}
\label{aca}
{\cal L}_0&=&R - {1\over 2 \phi_2^2} \partial_\mu S \partial^\mu {\bar S} - {1\over 12 \phi_2} (S H^1 + H^2)_{\mu\nu\rho} ({\bar S} H^1 + H^2 )^{\mu\nu\rho},\\
{\cal L}_3&= &{\zeta(3) \over 3. 2^6} e^{\frac{3}{2}\phi}(t_8 t_8 + \frac{1}{8} \epsilon_{10}\epsilon_{10}) R^4,
\end{eqnarray}
where $S=\phi_1 + i~ \phi_2= \chi + i e^{-\phi}$, $\zeta$ is the Riemann $\zeta$-function and we have used the compact notation

\begin{eqnarray}
(t_8 t_8 + \frac{1}{8} \epsilon_{10}\epsilon_{10}) R^4 &\equiv & 
 ({t_8}^{\mu_1 \mu_2 \ldots\mu_8} {t_8}^{\nu_1 \nu_2 \ldots\nu_8} + {1\over 8} {\epsilon_{10}}^{\mu_1\mu_2 \ldots\mu_8\mu_9\mu_{10}}  {{\epsilon_{10}}^{\nu_1 \nu_2 \cdots \nu_8}}_{\mu_9\mu_{10}})\times \nonumber\\
 & &{\hat R}_{\mu_1\mu_2\nu_1\nu_2}{\hat R}_{\mu_3\mu_4\nu_3\nu_4}{\hat R}_{\mu_5\mu_6\nu_5\nu_6} {\hat R}_{\mu_7\mu_8\nu_7\nu_8} +\ldots,
\end{eqnarray}
where $t_8$ is the standard eight-index tensor arising in string amplitudes,  $\epsilon_{10} $ is the totally antisymmetric symbol in ten dimensions and $\ldots$ denotes terms involving derivatives of the dilaton and 

\begin{equation}
\label{ad}
{{\hat  R}_{\mu\nu}}^{\alpha\beta}= {R_{\mu\nu}}^{\alpha\beta} - {1\over 4}g_{[\mu}^{[\alpha} \nabla_{\nu]}\nabla^{\beta]}\phi .
\end{equation}

   Lagrangian  (\ref{a}) reproduces the four-point amplitude calculated in string theory and it is in agreement with  $\sigma$-model calculations. Furthermore, ${\cal L}_0$ is invariant under an  SL(2,R) symmetry that acts as

\begin{equation}
\label{ab}
g_{\mu\nu}\rightarrow g_{\mu\nu},\quad  S \rightarrow {aS + b \over c S + d},\quad  \left(\begin{array}{cc} a & b\\ c & d  \end{array}\right)\in SL(2,R)\ 
\end{equation}
A discrete subgroup of this symmetry, $SL(2,Z)$, is conjectured to be an exact non-perturbative symmetry of the type IIB string \cite{schwarz}.

At the perturbative level, there exist string one-loop corrections to the four-point functions. For four gravitons, these corrections have been calculated and amount to the exchange \cite{sakai}

\begin{equation}
\label{ah}
\zeta(3)\rightarrow \zeta(3)+{\pi^2 \over 3} \phi_2^{-2}
\end{equation}
in Eq.~(\ref{aca}). Non-renormalization theorems ensure that there are no higher-loop corrections \cite{tseytlin}.
On the other hand, in type IIB theory there can be non-perturbative contributions due to  D-instantons \cite{green}. The multi-instanton contributions are determined by the SL(2, Z) symmetry i.e. the exact non-perturbative result for four gravitons can be found  by covariantizing the perturbative result, which is not SL(2, Z)-invariant, under SL(2, Z) \cite{green}

\begin{equation}
\label{ai}
{\cal L}_{3}={1\over 3.2^7} f_0(S, \bar S) \left(t_8 t_8 + {1\over 8} \epsilon_{10} \epsilon_{10} \right) R^4,
\end{equation}
where we have used the compact notation defined in  Eq.~(\ref{aca}); $f_0(S,{\bar S})$ is the non-holomorphic modular form 
\begin{equation}
\label{aj}
f_0(S, {\bar S})=\sum_{(m,n)\not=(0,0)}{\phi_2^{3/2} \over \vert m + n S\vert ^3}
\end{equation}
and  has the small $\phi_2$ expansion

\begin{equation}
\label{al}
f_0=2\zeta(3) \phi_2^{3/2} + {2 \pi\over 3} \phi_2^{-1/2} + 8 \pi \phi_2^{1/2} \sum_{m=0, n\geq 1}\left\vert{ m \over n  }\right\vert e^{2 i \pi m n \phi_1} K_1(2\pi \vert m n\vert \phi_2),
\end{equation}
where $K_1 $ is a Bessel function.
This form  for the exact $R^4$ corrections satisfies SL(2,Z) invariance, reproduces the correct perturbative expansion and the non-perturbative corrections are of the expected form.

We  look for  maximally symmetric solutions of  ${\cal L}_0 +  {\cal L}_3 $ (provisionally we set $\chi=0$).  For  maximally symmetric spaces

\begin{equation}
\label{af}
 R^\mu_{\phantom{\mu} \nu\lambda\sigma}=K\left(
\delta^\mu_\lambda g_{\nu\sigma} - \delta^\mu\sigma g_{\nu\lambda}\right), \qquad \phi=\phi_c=cte.
\end{equation}

It is known \cite{bento1} that it is not possible to satisfy both the graviton and dilaton equations of motion with ${\cal L}_3$ given by Eq.~(\ref{aca}), except for $K=0$ i.e. Minkowsky space is the only maximally symmetric solution possible. However, if we take ${\cal L}_3$ given by  Eq.~(\ref{ai}), the equations of motion for the metric and scalar fields read, respectivey \cite{bento3}

\begin{eqnarray}
\label{am} 
\alpha_1 K + { \alpha_2} f_0(S, {\bar S}) K^4 & = &0,\\
{ \alpha_2} f_0(S, {\bar S})_{,S} K^4 & = & 0,\\
{ \alpha_2} f_0(S, {\bar S})_{,{\bar S}} K^4 & = & 0,
\label{an}
\label{ana}
\end{eqnarray}
 where $\alpha_1=(D-1)(D-2)$ and $\beta=- {3\over 8}\zeta(3) e^{-3/2 \phi} (D-3)(D-2)(D-1) $and ${\alpha_2}=- {3\over 2^4}(D-3)(D-2)(D-1)$.

  Hence, it is clear that it is now possible to satisfy all  equations provided  extrema of $f_0$ exist. This is the case as the fixed points, $S=e^{i \pi/6}$ and $S=1$ in the fundamental domain, are necessarily extrema of $f_0$. We have checked numerically that, in either case,  $f_0 > 0$  and, therefore, substituting in eq.~(\ref{am}), we find $K_s=(-{\alpha_1 \over f_0{\alpha_2}})^{1/3}>0$, corresponding to a de Sitter solution. However, this solution is unstable  since, defining ${\bar f}(K)
= \alpha_1 K + f_0{\alpha_2} K^4 $, then ${\bar f}'(K_s)=-3\alpha_1 < 0$. There remains, of course, the possibility that stable solutions become possible when higher order curvatures are included.

For type IIA theories, the tree level effective Lagrangian  for $R^4$ terms is the same as for type IIB theories and   $R^4$ terms  receive no perturbative corrections beyond one loop; however, for type IIA theories there are  no non-perturbative corrections  in ten-dimensional type IIA theories  \cite{green2}. Hence, given the analysis presented above, it is not possible to obtain de Sitter (anti-de Sitter) solutions for these theories.

Finally, we turn to the question of whether the de Sitter solutions we found for type IIB theories up to order $\alpha'^3$ survive when higher order curvatures are taken into account. One expects that the  equations of motion (\ref{am})--(\ref{ana})  generalise to 

\begin{eqnarray}
\alpha_1 K + \alpha_2\ f_0(S, {\bar S})\ K^4 + \ldots + \alpha_{n-2}\ f_{n}(S, {\bar S})\ K^{n+4} + \ldots &=&0,\\
f_0(S, {\bar S})_{,S}\ \alpha_2\ K^4 + \ldots + \alpha_{n-2}\ f_n(S, {\bar S})_{,S}\ K^{n+4} + \ldots&=&0,\\
f_0(S, {\bar S})_{,\bar S}\ \alpha_2\ K^4 + \ldots + \alpha_{n-2}\ f_{n,\bar S}(S, {\bar S})_{,\bar S}\ K^{n+4} + \ldots&=&0.
\label{ba}
\end{eqnarray}

 Hence, it is clear  that it is possible to satisfy all  equations if the functions $f_n$  have at least one common extremum. Normally this would not happen but, for the case of type IIB superstrings, the SL(2,Z) symmetry requires  the functions $f_n$ to be  SL(2,Z)-invariant, implying that  they have common  fixed points, $S=e^{i \pi/6}$ and $S=1$, and these are then necessarily extrema of these functions \cite{shapere}. Hence, we conclude that maximally symmetric solutions exist, in principle,  for the superstring II  at all orders in $\alpha'$.
 Although this result is quite general and does  not depend on the particular form of the functions $f_n$  but solely on the assumption of SL(2,Z)-invariance, it is relevant  to mention  that the structure of  the higher order curvature terms has  already been discussed in the literature and a proposal exists for their form \cite{partouche}

\begin{equation}
\int d^{10} x \sqrt{-g} f_{-\frac{m}{2},0}(S,\bar S)\ R^{3m+1},
\end{equation}
where 

\begin{equation}
f_{s,k}(S, \bar S)=\sum_{(m,n)\not=(0,0)}{\phi_2^s\over (m S +n)^{s+k} (m\bar S + n)^{s-k}}
\end{equation}
are non-holomorphic modular forms of weights $(k, -k)$; hence, the functions $f_{-\frac{m}{2},0}$ are indeed modular invariant. Notice that $f_{3/2,0}$ is identical to $f_0$ introduced before.

 Hence, we see that the existence of de Sitter solutions for type IIB superstrings seems to be intimately related to the SL(2,Z) symmetry of the theory. In this context, it is interesting to notice  that, although stabilization of the dilaton and sucessfull inflation are very difficult features to implement for generic string cosmological models \cite{binetruy},  the situation seems to improve significantly for S-field  potentials that are  based on the assumption of SL(2,Z) invariance in the context of $N=1$ supergravity \cite{bento2}.

We conclude that inclusion of non-perturbative effects in type IIB superstrings
make it possible to obtain de Sitter solutions to the effective action at all orders in  $\alpha'$.

\section*{Acknowledgments}

We are grateful to A. Kehagias, O. Bertolami and especially to F. Quevedo for valuable discussions.


\end{document}